\documentclass[twocolumn,prb]{revtex4}
\usepackage{amsmath,amsthm,amscd}
\usepackage{dcolumn}

\usepackage{graphicx}
\usepackage{epsfig}
\begin{document}
J. Stat. Phys., to appear.

\title{The Chemical Potential}
\author{T. A. Kaplan}
\affiliation{Department of Physics and Astronomy and Institute for Quantum Sciences, \\ Michigan State University,
East Lansing, Michigan
48824}

\begin{abstract}
The definition of the fundamental quantity, the chemical potential, is badly confused in the literature: there
are at least three distinct definitions in various books and papers. While they all give the same result in the
thermodynamic limit, major differences between them can occur for finite systems, in anomalous cases even for
finite systems as large as a cm$^3$. We resolve the situation by arguing that the chemical potential defined as
the symbol $\mu$ conventionally appearing in the grand canonical density operator is the uniquely correct
definition valid for all finite systems, the grand canonical ensemble being the only one of the various
ensembles usually discussed (microcanonical, canonical, Gibbs, grand canonical) that is appropriate for
statistical thermodynamics, whenever the chemical potential is physically relevant. The zero-temperature limit
of this $\mu$ was derived by Perdew et al. for finite systems involving electrons, generally allowing for
electron-electron interactions; we extend this derivation and, for semiconductors, we also consider the zero-T
limit taken after the thermodynamic limit. The enormous finite size corrections (in macroscopic samples, e.g. 1
cm$^3$) for one rather common definition of the c.p., found recently by Shegelski within the standard effective
mass model of an ideal intrinsic semiconductor,
are discussed. Also, two very-small-system examples are given, including a quantum dot.
\end{abstract}

\maketitle
\section{Introduction}
This paper deals with the conceptual problem of defining the fundamental quantity, the chemical potential, in
statistical mechanics. Understanding of this matter is important in many places in physics and chemistry. That the
account in the present literature is badly confused will become clear just below. To gain this understanding, I
have found that the relation between equilibrium statistical mechanics and thermodynamics has to be reconsidered.

Two distinct issues are discussed here. One is the definition of the chemical potential, which I will call c.p.
until the question is resolved. The other is the question of the accuracy of the thermodynamic limit for a piece
of a semiconductor as large as a cm$^3$. We begin with a specific example, an ideal semiconductor, that serves
to introduce both issues; later, other quite different, models and related experiments are used to illustrate
the issue of definition in connection with very small systems.

Recently, Shegelski~\cite{shegelski1, shegelski2} defined the quantity
\begin{equation}
\phi(N)\equiv F(N+1)-F(N),\label{H.1}
\end{equation}
as being the c.p. of an $N$-electron system. Here $F$ is the Helmholtz free energy, as calculated in the
canonical ensemble, and the difference is taken at constant temperature $T$ and volume $V$.~\cite{footnote1}
Within the usual non-interacting-electron model (NEM) of an intrinsic semiconductor with $N$
electrons~\cite{kittel,ashcroftmermin,footnote2}, he found the following: The behavior of $\phi(N)$ as a
function of $T$ differs appreciably, at moderately low $T$, from the usual result~\cite{kittel,ashcroftmermin}
for the chemical potential, \textbf{even for systems with volumes the size of the universe!}, although in the
thermodynamic limit ($V,N\rightarrow\infty, N/V$ fixed) $\phi(N)\rightarrow$ the usual result for all $T>0$. He
finds $\phi(N)=\epsilon_c$ at $T=0$ for any finite volume $V$, where $\epsilon_c$ is the energy at the bottom of
the conduction band. In contrast, the thermodynamic limit (TL) of $\phi(N)\rightarrow (\epsilon_c+\epsilon_v)/2$
as $T\rightarrow0$, where $\epsilon_v$ is the energy at the top of the valence band; i.e. the thermodynamic
limit of $\phi(N)$ approaches the middle of the band gap, as does the usual result. Shegelski's result behaves
qualitatively exactly as one would expect without calculation: At $T=0, F(N)=E(N)$, the ground state energy for
$N$ particles, so that for the model considered $\phi(N)=\epsilon_c$, as easily shown from the expression for
$E(N)$. And at $T>0$ (see~\cite{LL} and Appendix B in \cite{ashcroftmermin})\cite{footnoteTL}, in the TL
\begin{equation}
\phi(N) \rightarrow \frac{\partial (F/V)}{\partial (N/V)}=\frac{\partial F}{\partial N}, \label{H.2}
\end{equation}
the derivative taken at constant $T$. ($F/V$ is the limiting function~\cite{LL}.) Thus this is the usual result,
approaching the middle of the gap as $T\rightarrow0$. The equality is formal, presented to exhibit the last form
as the usual definition of the c.p. in thermodynamics~\cite{callen}, where $N$ is assumed to be continuous. What
is surprising and most interesting are the enormous finite-size corrections. Such a scenario where the TL cannot
be essentially reached with a cm$^3$ sample is certainly discomforting, but in light of Shegelski's results,
clearly must be considered, as done below.

These results force extra care with terminology. I will use the term `macroscopic' to mean systems with roughly
Avogadro's number of particles, to within a few factors of 10 either way, to be contrasted with the TL. I will
mean by `finite systems' those with the number of particles ranging from a few (say 1,2,...100) up to
macroscopic size. I will also use the term `small systems' to denote those with a few particles.

First I note that the identification of $\phi$ with the c.p. is found elsewhere, e.g in the book by Ashcroft and
Mermin.~\cite{ashcroftmermin} Landsberg and Browne~\cite{landsberg} also apparently accept $\phi$ as the c.p.: in
reference to Shegelski~\cite{shegelski1} they state ``it is confirmed that the Fermi level of an intrinsic
semiconductor goes to the bottom of the conduction band as $T\rightarrow 0$." But they note that ``however the
loss of one electron is sufficient to make it go to the top of the valence band, and for some purposes it may be
adequate  to take an average of the two values". Evidently they are identifying the c.p. with the Fermi level, and
they seem to be saying that the c.p. is not a uniquely defined quantity.

Another definition is given in the book by Kittel and Kroemer~\cite{kittel2}, namely the c.p. for an $N$-particle
system is identified with
\begin{equation}
\psi(N)\equiv F(N)-F(N-1),\label{H.3}
\end{equation}
this being the top of the valence band at $T=0$ in the NEM. Fetter and Walecka~\cite{fetter} give both~(\ref{H.1})
and~(\ref{H.3}) at $T=0$ (they confine themselves to the TL, but as we have seen, these differ for our model
semiconductor, even in the TL). Hill~\cite{hill}, considering general $T$ and small systems, gives~(\ref{H.1}) but
notes that~(\ref{H.3}) would be just as good a choice. Chaikin and Lubensky~\cite{chaiken} define the chemical
potential, in the context of thermodynamics, as the change in internal energy produced by addition of one
particle, apparently supporting (\ref{H.1}). Baierlein~\cite{baierlein} gives c.p. $=\frac{\partial F}{\partial
N}= \psi(N).$

Perdew et al~\cite{perdew,perdew2} showed for systems with $N_0$ electrons that the zero-$T$ limit of the quantity
$\nu$ that appears in the grand canonical density operator,
\begin{equation}
 \rho=exp[-\beta(H-\nu N)]/tr\ exp[-\beta(H-\nu N)],\label{H.4}
\end{equation}
is
\begin{equation}
\nu(N_0)=\frac{1}{2}[E(N_0+1)-E(N_0-1)],\label{H.5}
\end{equation}
where $\beta=1/(kT)$, and $N_0$ is the average value of the number of particles, taken to be
integral\cite{footnoteperdew}. See also~\cite{kaplan7}. Since $\nu$ is also quite generally called the chemical
potential, we clearly have another definition of the c.p. It is the average of $\phi$ and $\psi$ at $T=0$,
mentioned in~\cite{landsberg}.

For metals, the various formulas give essentially the same result for $N \sim 10^{23}$. It is for small systems,
(e.g. a Na atom or a quantum dot) and for semiconductors that one needs to be concerned about the difference
between them. It is well known (see e.g. ref.~\cite{schrodinger}) that in the TL, within the NEM, the average
occupation number of a 1-electron state with energy $\epsilon$ is the Fermi-Dirac distribution
\begin{equation}
<n_\epsilon> = f(\epsilon-\nu)\equiv[exp\beta(\epsilon-\nu)+1]^{-1}.\label{H.6}
\end{equation}
In the well known and very simple derivation of (\ref{H.6}) via the grand canonical ensemble, \emph{the quantity
$\nu$ is the same as that appearing in} (\ref{H.4}), (\ref{H.6}) \emph{holding independently of system size}. It
is also well known that, as we have said, in the thermodynamic limit taken at $T>0$, within the NEM, $\nu$, (the
quantity usually actually calculated as the c.p. in the NEM), approaches (\ref{H.5}), which is the middle of the
gap for the semiconductor model, as $T\rightarrow 0$  (see, e.g.~\cite{kittel,ashcroftmermin}). But we need to
be concerned about the role of finite size, as we said.

Note that (\ref{H.5}) can be written
\begin{equation}
\nu(N_0)=-\frac{1}{2}[I(N_0)+A(N_0)],\label{H.7}
\end{equation}
where
\begin{eqnarray}
I(N_0)&=&E(N_0-1)-E(N_0)\nonumber\\
A(N_0)&=&E(N_0)-E(N_0+1)\label{H.8}
\end{eqnarray}
are respectively the ionization energy and electron affinity of the $N_0$-electron system at $T=0$. Parr et
al~\cite{parr} note that this is closely related to a formula due to Mulliken~\cite{mulliken1}, who spoke of the
electronegativity (e.n.) rather than the chemical potential, and gave a formula for e.n. as the negative of
(\ref{H.7}). See also Gyftopoulos and Haftopoulos\cite{gyftopoulos} who simply define e.n. as -$\nu$ (arguing that
``since $\nu$ is interpreted as the escaping tendency (the opposite of the power to attract) of a component  from
a thermodynamic system, it is reasonable to use its negative as a measure of electronegativity"), and
~\cite{pauling2, perdew,perdew2} for related discussion. It appears that the analysis in~\cite{mulliken1} is
approximate, involving tight-binding molecular orbitals and Hartree-Fock approximation; whereas (\ref{H.5}) was
proved more generally (also see below).

Thus we have several distinct definitions for the chemical potential appearing in the literature. For
mathematics only, one can take his/her pick. However, there is an important piece of physics that removes this
arbitrariness, in my opinion. Namely, the condition for equilibrium with respect to particle flow between two
systems at the same temperature is that their chemical potentials be equal, the chemical potentials in this
context being defined as the values of $\nu$ appearing in their respective distributions (\ref{H.4}) (see e.g.
Tolman~\cite{tolman}). Thus, thinking in terms of experiment, equilibrium, i.e. zero current, will occur e.g.
between a metal and a semiconductor in contact and at the same $T$ when $\nu$ for the semiconductor equals that
in the metal. In standard thermodynamics, it is the relation~(\ref{H.2}), $c.p.=\partial F/\partial N$, plus
minimization of the total free energy, that leads to this fundamental equilibrium condition.~\cite{callen}

In Section II, I argue that the grand canonical ensemble is the proper basis for a theory of statistical
thermodynamics, and correspondingly, it is $\nu$ that is uniquely the appropriate meaning of the c.p. The
argument is supported by two very-small-system examples, a quantum dot and a pair of atoms contemplating charge
transfer.

In Section III, I discuss the zero-T limit of $\nu$, both for finite systems, and, in the case of solids,
whether the limit is taken before or after the TL. Section IV describes a simple calculation of $\nu$ for a
finite semiconductor within the same NEM discussed above, with attention to the approach to the TL. It is shown
that this quantity does not show the strange huge finite-size effects found for $\phi$, $\nu$ being very close
to the TL  for macroscopic size, at all $T$. In the Appendix I give a somewhat simpler calculation than
Shegelski's which supports his finding of a huge finite-size effect (this shows up for $\nu(N_0+1), N_0$
corresponding to the neutral semiconductor).

As in the relevant references cited~\cite{shegelski1,shegelski2,kittel,ashcroftmermin}$^{,}$
\cite{landsberg}$^{-}$\cite{fetter}$^{,}$\cite{perdew,perdew2}, we will be considering the very common
one-component model where the electrons move in a fixed one-electron potential. However, the considerations of
Sections II and III, with the exception in the latter of the quantum dot model, are more general: They allow for
the adjustment of the one-electron potential as one adds or removes electrons. E.g., in the hydrogen molecule,
the equilibrium separation of the protons will differ for the neutral and for the charged cases, such
adjustments being allowed for in the formalism; a given molecule in the usual chemical sense has a given number
of nuclei with different degrees of ionization. Alternately, when we consider the thermodynamic limit of a
solid, we add nuclei as we add electrons in the customary way: in calculating the TL of $F(N)$ for the
semiconductor, the net charge is kept constant (at 0); similarly, for $F(N+1)$, the net charge is again held
constant (at 1 electronic charge). Also, the generalization to multi-component systems of the final choice of
definition of the c.p. is mentioned in the next section.

I mention here that in Section IV (Final Remarks), I emphasize the fact that the ideal intrinsic semiconductor
is quite special (anomalous), and although it is clearly of considerable pedagogical importance, the
very-small-system examples are of more physical interest.

\section{Reconsideration of statistical thermodynamics}
The usual definition of $F$ is via the canonical ensemble:
\begin{eqnarray}
F(N)&=&-kT \mbox{ln} Z_c(N)\label{H.9}\\
 Z_c(N)&=&\sum_j \mbox{exp}[-\beta E_j(N)],\label{H.10}
\end{eqnarray}
where $j=1,2,\cdots$ and $E_1(N)\le E_2(N)\le\cdots$ are the energy eigenvalues for states of $N$ particles. We
have assumed that the Hamiltonian and the operator $\hat{N}$ for the number of particles commute. $N$ is of course
discrete, so that differentiation with respect to $N$ is not defined. If an analytic expression for $F(N)$ were
known, then, for large $N$, continuing the function to continuous $N$ would make sense; e.g. if $F(N)=N+ln N$,
then $\phi(N)=1+ ln (1+1/N)$, which approaches $dF/dN$ for large $N$. But generally such knowledge is not
available; while interpolating in some smooth way between the integer values might be satisfactory, this entails
some loss of rigor, and might be troublesome, particularly for small systems. Surely this has been the motivation
behind \emph{defining} the chemical potential in terms of finite differences of $F(N)$ as in (\ref{H.1}) and
(\ref{H.3}). A similar situation exists for the other commonly discussed distributions, microcanonical, or the
Gibbs distribution (fixed $T$ and pressure), with the exception of the grand canonical ensemble. In the latter
case, $N$ is merely a summation variable, the role of the number of particles is played by the average value of
$N$,
\begin{equation}
\mathcal{N}=\frac{\sum_{j,N}N e^{-\beta[E_j(N)-\nu N]}}{\sum_{j,N}e^{-\beta[E_j(N)-\nu N}]},\label{H.11}
\end{equation}
which is a continuous variable at $T>0$.

In standard statistical thermodynamics, each of the various distributions gives rise naturally to a
corresponding thermodynamic potential or free energy. E.g., the canonical ensemble gives directly the Helmholtz
free energy: defining the entropy as $S=-k $tr$\rho_c \mbox{ln}\rho_c$ where $\rho_c=$exp$(-\beta H)/Z_c$, and
the trace is taken over states with fixed number of particles $N$, one sees easily that $F=U-TS$ ($U$ = average
energy). And $F$ comes out directly as a function of $T, V, N$. In thermodynamic theory~\cite{callen}, knowledge
of $F(T,V,N)$ as a function of the variables indicated, the natural variables~\cite{callen} for $F$, is a
fundamental equation; that is, from it one can calculate all thermodynamic properties, including other free
energies \emph{and the chemical potential}. Similarly, the grand canonical ensemble gives the grand canonical
free energy, called the thermodynamic potential by Landau and Lifschitz~\cite{landau2},
\begin{eqnarray}
\Omega&=&-kT \mbox{ln}\ Z.\label{H.12}\\
Z&=& \mbox{tr exp}[-\beta(H-\nu \hat{N})]\label{H.13}
\end{eqnarray}
is the grand canonical partition function, which is given directly in terms of the variables $T,V,\nu$, the
natural variables for $\Omega$. (The trace here differs from that for the canonical ensemble in that here it
involves summing over energy states with various numbers of particles.) In this case the entropy is
$\mathcal{S}=-k$ tr $\rho$ ln $\rho$ where $\rho$ is the grand canonical density operator (\ref{H.4}). Using this
we see that
\begin{eqnarray}
\Omega&=&\mathcal{U}-T\mathcal{S}-\nu\mathcal{N},\label{H.14}\\
&=&\mathcal{F}-\nu\mathcal{N}\label{H.15}
\end{eqnarray}
where
\begin{equation}
\mathcal{U}=\frac{\sum_{j,N}E_j(N) e^{-\beta[E_j(N)-\nu N]}}{\sum_{j,N} e^{-\beta[E_j(N)-\nu N]}}. \label{H.16}
\end{equation}
The change in notation, e.g. $U\rightarrow\mathcal{U}$, is made to distinguish between quantities calculated in
the canonical and grand canonical ensembles. In the thermodynamic limit, such a distinction is unnecessary, but,
again, we have to consider finite size effects. From the mathematics of thermodynamic theory, knowing $\Omega$ as
a function of $T,V,\nu$, we can construct all thermodynamic quantities. However, the corresponding Helmholtz free
energy $\mathcal{F}$, internal energy $\mathcal{U}$ and entropy $\mathcal{S}$ will differ from their corresponding
quantities $F, U, S$ by virtue of the different ensembles being used in the two cases. In particular,
$\mathcal{F}$ is calculated via (\ref{H.15}):
\begin{equation}
\mathcal{F}=\Omega+\nu\mathcal{N}.\label{H.17}
\end{equation}

The continuous nature of $\mathcal{N}$ allows differentiation of~(\ref{H.17}) with respect to $\mathcal{N}$; doing
this at constant $T$ and volume $V$, and recognizing the $\mathcal{N}$-dependence of $\nu$ generated
by~(\ref{H.11}), yields the identity
\begin{equation}
\frac{\partial \mathcal{F}}{\partial \mathcal{N}}=\nu,\label{thermomu}
\end{equation}
as is well known (see, e.g.~\cite{tolman,perdew2}). This is of course recognized as the historical thermodynamic
relation between the chemical potential and the Helmholtz free energy. But note that it is exact even for finite
systems; the TL is not required for its validity.

Now consider the \emph{\textbf{definition of the chemical potential}}. It is clear that choosing $\phi$ or $\psi$
would be entirely arbitrary, so neither is satisfactory. The average of these would be a guess, as would be either
of them; i.e., the only basis for such guesses (apparently) is that they reduce to the thermodynamic definition in
the TL, and of course there is an infinity of such guesses. The idea that the c.p. can be chosen in various ways
~\cite{landsberg,hill} is unacceptable, for reasons already stated. Thus, in light of~(\ref{thermomu}), and the
role of the chemical potential in particle transport as argued by Tolman, plus the rather universal understanding
of this role (see e.g. \cite{degroot, blatt, ferry}), I suggest that the proper definition of the chemical
potential is
\begin{equation}
c.p.= \nu \equiv\mu.\label{H.18}
\end{equation}
(I have now bestowed on this the symbol $\mu$, used almost universally for the chemical potential.) Given this
definition, most of the other standard thermodynamic relations follow, i.e. all that don't depend on the
assumption of extensivity of the ``extensive" variables, or in other words the assumption that
$\mathcal{U}(\mathcal{S},V,\mathcal{N})$ is a homogeneous first order function. E.g., it is well-known that, along
with~(\ref{H.14}) and~(\ref{thermomu}), the relations
\begin{eqnarray}
\left( \frac{\partial \Omega}{\partial T} \right)_{\mu,V}&=&-\mathcal{S}\nonumber\\
\left( \frac{\partial\Omega}{\partial \mu}\right)_{T,V}&=&-\mathcal{N}\nonumber\\
d\mathcal{U}&=&T d\mathcal{S}+\mu d\mathcal{N}-p dV,\nonumber
\end{eqnarray}
where
\begin{equation}
 p\equiv -\left( \frac{\partial \Omega}{\partial{V}} \right)_{T,\mu},\nonumber
\end{equation}
follow directly from the definitions given. But in general one does \emph{not} have the relations that follow from
the extensivity hypothesis\cite{callen}:
\begin{eqnarray}
 \mathcal{U}&\ne& T\mathcal{S}-p V + \mu \mathcal{N}\nonumber\\
 \Omega &\ne& -pV\nonumber\\
 G&\equiv&\mathcal{F}+pV\ne \mu N.\nonumber
 \end{eqnarray}
 Well, so be it--the physical information contained in the
density matrix relating to the average ``extensive" quantities is given only by the thermodynamic
\emph{eq}ualities listed plus other relations derivable from these. The inequalities listed are written by
Hill~\cite{hill} through the definitions $\hat{\mu}=G/N, \hat{p}=-\Omega/V$; then
\begin{equation}
\mathcal{U}=T\mathcal{S}-\frac{p+\hat{p}}{2}V + \frac{\mu+\hat\mu}{2}\mathcal{N}.\nonumber
\end{equation}
Of course $\hat{p}\rightarrow p,\ \hat{\mu}\rightarrow\mu$ in the TL.

Clearly the only statistical approach that allows the rigorous connection to thermodynamics,~(\ref{thermomu}),
is through the grand canonical ensemble (among the usually considered ensembles). It seems reasonable to
conclude that for finite systams \emph{the grand canonical ensemble is the only legitimate choice from the point
of view of statistical thermodynamics, whenever the chemical potential is relevant}. The view proposed is simply
that all of statistical thermodynamics where the chemical potential is physically relevant be carried out on the
basis of the grand canonical ensemble.\cite{footnotelocalspins,footnote3, cabib} This then obviates the apparent
necessity to define the c.p., \emph{ad hoc}, as some functional of a free energy defined only on the integers. I
will work with this new view and the definition~(\ref{H.18}) henceforth.\cite{footnotegeneralization}

The above argument deserves further scrutiny. Tolman's proof about equality of $\mu$ being the condition of zero
particle flow between two systems assumes that at thermal equilibrium the energy eigenvalues are the sums of the
eigenvalues of the two systems. This is essentially the same assumption made in regard to the similar question
regarding the relation between temperature and heat flow (see also~\cite{schrodinger}). This assumption of
additivity was justified because the individual systems being considered were macroscopic. However, the
assumption can be questioned for finite systems, or where finite-size ``corrections" are being investigated, as
is our main interest here. In this connection, it has been shown that under the appropriate conditions, the
neglect of the interaction between a reservoir and a small system, is justified for the purpose of considering
thermal equilibrium.~\cite{cohen-tannoudji} The appropriate conditions are physically sensible requirements on
the reservoir (beyond being large), and small enough reservoir-system interactions. These lead at equilibrium to
a well-defined temperature $T$ of the small system, this $T$, along with the concept of heat flow, having
typical attributes of temperature for large systems, in the sense of energy exchange between the small system
and the reservoir.

We expect (and assume) that a similar result will hold for the chemical potential and particle flow. In support of
that expectation, we consider two examples based in physical reality, (a) the case of quantum dots, and (b) the
prediction of ionicity in diatomic molecules (and some related solids).

(a) Experiments on quantum dots indicate strongly that under appropriate conditions, one can ascribe with high
certainty whether there is one or two, etc. electrons on the dot.~\cite{tarucha,ashoori,ashoori2} One might
worry that description by a grand canonical ensemble, which gives fluctuations in the number of particles, which
might be large for small systems, would therefore contradict these experimental results. Further, while I
believe the physics of these experiments is well understood, there is confusion about the concept of the
chemical potential of the dot, at least in some writings. The following extremely simple example shows that the
grand canonical ensemble can answer the question about the fluctuations, and illustrates how a proper
understanding of the chemical potential gives a picture consistent with experiments. We consider the single-site
Hubbard model,
\begin{equation}
H=-I_1 n + Un_\uparrow n_\downarrow,\label{hubbard}
\end{equation}
where $U (> 0)$ is the electron-electron Coulomb interaction, $n_\sigma$ is the occupation number for a
spin-$\sigma$ electron, and $n=n_\uparrow+n_\downarrow $. $I_1 > 0$ is the ionization potential when there is 1
electron. This was used in previous considerations of a pair of quantum dots (plus an inter-dot hopping
term).~\cite{hu,schliemann,kaplan} Recognizing that 2$n_\uparrow n_\downarrow = n^2-n$, one sees that it is
essentially the capacitive term also used to represent the so-called ``charging energy" in many single-dot
models.~\cite{kastner,kouwenhoven} There are only 4 states in the model, with just 3 values of $n$, namely, 0,1,2.

\begin{figure}[h]
\centering\includegraphics[height=120pt]{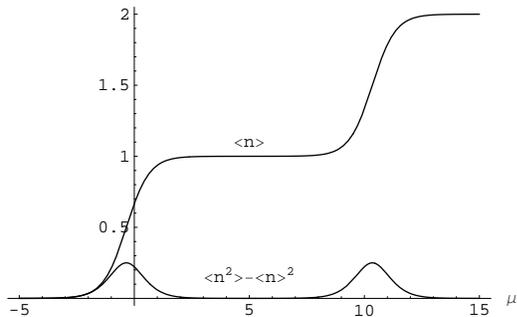}
 \caption{Average occupation number $<n>$ and its mean square fluctuations vs. $\mu$ in meV. ($U=10$ meV.)}
 \label{fig:naverage}
\end{figure}

The grand canonical distribution readily yields the (grand) partition function $Z=1+2e^{\beta
\mu}+e^{2\beta\mu-\beta U}$, the average number of particles $<n>$, and the mean square fluctuation
$<n^2>-<n>^2=kT\frac{\partial<n>}{\partial\mu}$, shown in FIG.~\ref{fig:naverage} for the typical
values~\cite{kaplan,tessmer} $U=10$meV, $T=U/20=6K$; $I_1$ has been put to 0, amounting merely to a shift in the
zero of $\mu$ (from -$I_1$). It is seen
that plateaus occur at integer values of $<n>$, and for $\mu$ in the corresponding range, the fluctuations in
$n$ are negligible. Thus small fluctuations have been achieved because the temperature is low compared to $U$,
rather than for the usual reason, large system size. The plateaus in $<n>$ illustrate the Coulomb blockade,
commonly discussed in connection with transport experiments,~\cite{tarucha,kastner,kouwenhoven,steffens}, but
see~\cite{ashoori,ashoori2}. (If $U=0, <n>$ would rise directly to the value 2, the plateau at $<n>=1$ not
occurring.) This effect is already well-known in atoms, it being the result of the Coulomb repulsion, manifested
in the fact that the ionization potential exceeds the electron affinity (a picture for the lithium atom quite
similar to $<n>$ in Fig.~\ref{fig:naverage} appears in~\cite{perdew2}). The present discussion is strictly
within thermodynamic equilibrium, the relation with transport measurements being that the conductance of the dot
is small in the plateau regions, and large in the transition regions.

\begin{figure}[h]
\centering\includegraphics[height=100pt]{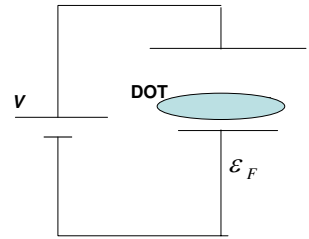}
 \caption{Schematic for capacitance spectroscopy on a quantum dot. Color online.}
 \label{fig:dotpicture}
\end{figure}

The basic idea of the experiments can be understood in some detail on the basis of this simple model, provided
we recognize the fundamental physical meaning of the chemical potential regarding electron flow, as discussed
above. One of the types of experiments, called single-electron capacitance spectroscopy~\cite{ashoori}, is
essentially an equilibrium experiment. It can be idealized by a system consisting of a metal lead connected to
the dot through a tunnelling barrier, as indicated schematically in FIG.~\ref{fig:dotpicture}. The proximity of
the dot to the lower lead (the reservoir) is meant to indicate that the barrier between the dot and that lead is
small, the tunnelling time $\tau$ being short, the barrier to the upper lead being very large, so tunnelling
through that is negligible.~\cite{dotdescription} The potential of the dot relative to the lead is controlled by
a gate voltage $V$, which has a DC and a small AC component. The energy of electrons on the dot due to this
potential is $-eV_d n$, $V_d$ being the potential on the dot, proportional to $V$. (We assume negligible
potential variation over the dot. Also, $e$ = magnitude of the electron charge.) On adding this term to the
Hamiltonian~(\ref{hubbard}), one readily sees that $<n>$ is a function of $\mu$ and $eV_d$ only through the sum:
$<n>=g(\mu+eV_d)$. Consider first the case where there is only a DC component. For a fixed $V$ the system will
come to equilibrium, so that $\mu=\epsilon_F$, the Fermi energy of the lead. For small $V$, $\epsilon_F$ will
deviate from its zero-$V$ value $\epsilon_F^0$ by a term linear in $V$, so that $<n>=g(\epsilon_F^0+cV)$, $c$ =
constant. Hence, by varying $V$ slowly (rate $<<1/\tau$, the equilibration rate), one can scan through the
abscissa of FIG.~\ref{fig:naverage}. The AC component is introduced to help detect the signal; its frequency is
small compared to $1/\tau$, so that at each instant the system can still be considered as essentially being in
thermodynamic equilibrium. Hence $\mu = \epsilon_F$ even in this case. Through this, the experiment directly
measures the charging of the dot, i.e. $\frac{dQ}{dV}$ (the capacitance of the dot, whence the terminology
capacitance spectroscopy), $\propto \frac{\partial<n>}{\partial\mu}$.~\cite{tessmer} This derivative is seen
from Fig.~\ref{fig:naverage} to consist of sharp peaks at the points where an electron is added to the dot, as
observed and interpreted in ref.~\cite{ashoori,ashoori2}. Since the derivative is proportional to the
fluctuations in $n$, the peaks in these experiments are directly proportional to those seen in the lower curve,
Fig.~\ref{fig:naverage}, although in the experiment, $T$ is considerably lower than assumed for the figure, so
the observed peaks are much sharper. (Also, the experiment considers the number of electrons going up to
remarkably larger values than 2, because the dot there has many more bound states than the model taken here for
simplicity.)~\cite{footnotebroadening}

In references~\cite{kouwenhoven,steffens} one finds the chemical potential of the dot defined at $T=0$ as
$\mu_{dot}(N)=\psi(N)$ ((\ref{H.3}) above) and $\mu_N=\phi(N)$ ((\ref{H.1}) above), respectively. These
definitions present a problem, seen as follows. For our model~(\ref{hubbard}) with $I_1=0$, $\phi(1)=U$. At $U$
on the horizontal axis, $<n>$ is either undefined (at $T=0$), or 3/4 (in the limit $T\rightarrow0$, as seen in
the next section). But presumably $\mu_1\equiv\phi(1)$ is supposed to be the chemical potential for particle
number $=1$, and hence it is either undefined or self inconsistent. A similar argument holds for the other
definition. In fact, at $T=0$, $\phi(N_0)\equiv -A(N_0)=\lim_{T\rightarrow0}\mu(\mathcal{N})$, for
$N_0<\mathcal{N}<N_0+1$, where $<n>\equiv\mathcal{N}$ and $N_0$ is an integer (see~(\ref{H.26})). Furthermore,
even in the zero-$T$ limit, there is not just one possible value of the chemical potential, as apparently
implied by each of these works; (\ref{H.26}) shows that there are three possible values of $\mu(\mathcal{N})$
for $\mathcal{N}$ in the range $N_0-1< \mathcal{N}<N_0+1$, two for non-integer $\mathcal{N}$, one for integer
$\mathcal{N} (=N_0)$ . (This is clear from Fig.~\ref{fig:naverage} for $N_0=1$ if one imagines the step-function
limit of $<n>$ as $T\rightarrow0$.)

(b) We consider the example discussed in~\cite{perdew,perdew2}, namely the question of charge transfer in typical
``ionic" diatomic molecules, but with a bit more emphasis on the role of the chemical potential. If our molecule
is BC, then at very large separation $R$ we know empirically that for minimum energy, atoms B and C are neutral
(since either $I$ is $>$ either $A$ for neutral atoms). As $R$ decreases, one considers the possibility of charge
transfer. The direction of this transfer, if it occurs, is determined of course by which direction gives minimum
energy; but I want to see how this depends on the chemical potentials of the atoms. Let the numbers of electrons
on the respective neutral atoms be $N_b,N_c$. The energies for the two directions of transfer are, for large $R$,
\begin{eqnarray}
E(B\rightarrow C)&=&E_b(N_b-1)+E_c(N_c+1)+v(R),\nonumber\\
E(B\leftarrow C)&=&E_b(N_b+1)+E_c(N_c-1)+v(R),\nonumber
\end{eqnarray}
where the leading term of the interaction energy $v(R)$ is $-e^2/R$, but can also contain higher polarization
terms; $E_i(N)$ is the energy of the isolated atom $i$. Thus, using~(\ref{H.5}),
\begin{equation}
E(B\rightarrow C)-E(B\leftarrow C)=2(\mu_C - \mu_B),\label{ionicity}
\end{equation}
showing that electron transfer from high to low chemical potential gives lower energy. This is of course
consistent with our requirement on a proper definition of the c.p.

\section{$T\rightarrow 0$ limit of $\mu$ and $\mathcal{N}$} As noted in the Introduction,
Perdew et al.~\cite{perdew,perdew2} derived (\ref{H.5}) for finite systems, allowing for exact treatment of
electron-electron interactions. Here we discuss the derivation, generalizing it, and in addition, consider, for
a semiconductor, taking the limit $T\rightarrow 0$ \emph{after} the thermodynamic limit (TL). It is well known
that for the non-interacting-electron model (NEM), the two limits commute for the semiconductor, i.e. when the
number of electrons exactly fills the valence band (see e.g. Section IV). That this however is not true in
general is rather obvious; it will be discussed below. We will also calculate the zero-T limit of $\mathcal{N}$.

 The model we consider is that of a number of electrons $N$ that move in a fixed (1-electron or external)
 potential, which can depend on $N$. As mentioned above, we are thinking of a molecule in a general sense (it can
 be a finite solid); the number of nuclei is fixed and their positions are fixed at positions that minimize the
 electronic energy in the Born-Oppenheimer sense. However, we will also at times consider the case where the
 potential is fixed, independent of $N$, a very common model for a solid; here we need to understand that in
 taking the thermodynamic limit, the usual procedure is to consider a sequence of such ``molecules", in each of which both
 the average number of electrons and the number of nuclei increase in proportion.

 It is of course known that there is a maximum number $N_{max}$ of electrons that will bind to a molecule
 ~\cite{gyftopoulos,lieb}, and clearly there is a minimum number $N_{min}$ such that the molecule (with a given
 number of nuclei) is bound. Defining the unbound states as describing a system other than the molecule, we
 consider only the bound states in the statistical mechanics, as done in~\cite{gyftopoulos} for atoms. Thus
\begin{equation}
\mathcal{N}=\frac{\sum_j\sum_{N=N_{min}}^{N_{max}} N exp[-\beta\Omega_j(N)]}{\sum exp[-\beta\Omega_j(N)]},
\label{H.19}
\end{equation}
where
\begin{equation}
\Omega_j(N)=E_j(N)-\mu N.\label{H.20}
\end{equation}
Clearly the $T\rightarrow0$ limit will be determined by the lowest ``free-energy eigenvalues", i.e. the lowest
values of $\Omega_j(N)$. For fixed potential, $N_{min}=0$. We will exclude the end-points $N_0=N_{min}, N_{max}$.
In those cases $\mu\rightarrow\pm\infty$.\cite{gyftopoulos}

An important assumption made is that the ground states satisfy
\begin{equation}
E_1(N-1)-E_1(N)> E_1(N)-E_1(N+1),\  N=1,2,3,\cdots\ .\label{H.21}
\end{equation}
This condition, which states that the ionization energy is $>$ electron affinity for any $N$, and which is a
convexity condition for the function $E_1$ defined on the integers, is usually true. It is expected intuitively,
but has not been proven, even for a system of electrons moving in the field of nuclei that are fixed in
position, as noted by Lieb~\cite{lieb2}. The intuition on which it is based is that it costs more energy to
remove a bound electron from an $N$-electron system ($I(N)$) than to remove it from an $N+1$-electron system
($A(N)$), since the leaving electron ``sees" a larger positive charge in the former case. However, if the nuclei
are allowed to relax from their positions in the $N$-electron system, both $E_1(N-1)$ and $E_1(N+1)$ will
decrease from their fixed-nuclei values, weakening the inequality; thus, even if~(\ref{H.21}) were true for
fixed nuclei, the inequality would not follow in general. Empirically it is true for all atoms; it is not known
to be violated for isolated molecules~\cite{harrison}. Also, it is easily seen to be true for all
non-interacting fermion models with fixed external potential, as well as the zero-bandwidth limit of the
repulsive-U Hubbard model (but not attractive U ~\cite{kaplan7}). But it is actively being questioned in
connection with experiments on quantum dots and similar systems. See e.g.~\cite{raikh}. We will content
ourselves with considering systems for which convexity holds.


As shown in~\cite{kaplan7}, the minimum of $\Omega_1(N)$ will occur for a given value $N_0$ of $N$,
given~(\ref{H.21}) and $\mu=\mu_0$, where
\begin{equation}
\mu_0 \equiv \frac{1}{2}[E_1(N_0+1)-E_1(N_0-1)].\label{H.22}
\end{equation}
Also, that analysis shows this minimum is unique, given~(\ref{H.21}). It is easy to see that for this $\mu$
\begin{eqnarray}
\Omega_1(N_0-1)-\Omega_1(N_0)&=&\Omega_1(N_0+1)-\Omega_1(N_0) \label{H.24}\\
&=&\frac{I(N_0)-A(N_0)}{2},\label{H.25}
\end{eqnarray}
where $I(N_0)$ and $A(N_0)$ are the ionization energy and electron affinity for the $N_0$-electron system.
(\ref{H.24}) says $\Omega_1(N)$ is symmetric about $N_0$ for the nearest neighboring values; such symmetry does
not occur for $|N-N_0|>1$ in general. One easily sees that the minimum of $\Omega(N)$ also will occur at and only
at $N_0$ for a range of $\mu$'s such that~\cite{footnotelandsbergtheorem}
\begin{equation}
-I(N_0) < \mu < -A(N_0),\label{H.27}
\end{equation}
$\mu_0$ lying in the middle of this range. However, at the end-points there is precisely 2-fold degeneracy:
\begin{eqnarray}
\Omega_1(N_0-1)&=&\Omega_1(N_0),\  \mbox{for}\  \mu=-I(N_0)\nonumber\\
\Omega_1(N_0+1)&=&\Omega_1(N_0),\  \mbox{for} \ \mu=-A(N_0).\label{H.25}
\end{eqnarray}
From this it is clear that \emph{at} $T=0$,
\begin{eqnarray}
\mu&=&-I(N_0)\Rightarrow  N_0-1 < \mathcal{N} < N_0,  \nonumber\\
\mu&=&-A(N_0)\Rightarrow N_0 < \mathcal{N} < N_0+1.\label{T=0}
\end{eqnarray}
This is the same as Eq. (6) of~\cite{perdew}. From this and~(\ref{H.19}) we can see further that
\begin{equation}
\lim_{T\rightarrow0}\mathcal{N}=\left\{ \begin{array}{ll}
N_0-\frac{1}{1+\gamma_-}  & \mbox{for}\  \mu=-I(N_0)\\
 N_0+\frac{1}{1+\gamma_+} & \mbox{for}\  \mu=-A(N_0),
 \end{array}
 \right.\label{borderline}
 \end{equation}
 where $\gamma_{\pm}=g(N_0)/g(N_0\pm1)$, $g(N)$ being the
 degeneracy of the ground level $E_1(N)$.
More generally, in~\cite{perdew2} it was argued that at low enough temperature, and for finite $N_0$, the three
terms in~(\ref{H.19}) coming from the ground states ($j=1)$ for $N=N_0-1,N_0,N_0+1$ dominate the sums for $N_0-1/2
< \mathcal{N} < N_0+1/2$.~\cite{footnotePPLB} In fact, this condition is overly restrictive, $N_0-1 < \mathcal{N}
< N_0+1$ being sufficient for the 3-state approximation. This approximation gives~\cite{perdew2}, with the larger
range of validity indicated,
\begin{equation}
\lim_{T\rightarrow0}\mu=\left\{ \begin{array}{ll}
-I(N_0),  & N_0-1 < \ \mathcal{N} < N_0\\
                                -\frac{1}{2}(I(N_0)+A(N_0)), & \mathcal{N}=N_0\\
                                -A(N_0), &   N_0 < \mathcal{N} < N_0+1
                                \end{array}
                                \right..\label{H.26}
\end{equation}
Note that the extended range in the 1st and 3rd cases includes~(\ref{T=0}) (excluded in~\cite{perdew2}),
important in the transition region between integers (see Fig.~\ref{fig:naverage}). The consistency of the
argument can be seen as follows. For non-integer $\mathcal{N}$, in the given range, (\ref{T=0}) says that $\mu$
must approach $-I(N_0)$ or $-A(N_0)$, yielding the degeneracy indicated in~(\ref{H.25}). In this case, these 2
degenerate states will dominate the sums, the 3rd state will be a finite energy higher, and therefore will not
contribute in the limit. In fact this ``2-state" approximation yields the first and third entries of eq. (52)
in~\cite{perdew2} (including the $T$-dependent term, the third state giving only an exponentially small
contribution). One needs the three states to simultaneously obtain the result~(\ref{H.26}) for both integer and
non-integer $\mathcal{N}$. For $\mathcal{N} = N_0$ (integer), the non-uniqueness of $\mu$ at $T=0$ seen
in~(\ref{H.27}) is removed at infinitesimal $T$, the value being uniquely that of~(\ref{H.5}) (or~(\ref{H.7})).
The restriction to finite $N_0$ is needed for the argument that only the ground levels of each $N$ enter at low
enough $T$.

As regards degeneracy of states with fixed $N$, Perdew's result~\cite{perdew2} shows that this only affects the
\emph{approach} to the limit~(\ref{H.26}). Incidentally, the linear-$T$ approach to the limit found in
\cite{perdew2} is also valid under the less restrictive conditions indicated in~(\ref{H.26}). It is interesting
to note that while the degeneracy does not affect the zero-T limit of $\mu$ (at fixed $\mathcal{N})$, it does
affect the limit of $\mathcal{N}$ (fixed $\mu$), as seen in~(\ref{borderline}). For the Hubbard model of the
previous section, $\gamma_\pm=2$, and Fig.~\ref{fig:naverage} shows that the limit is nearly achieved for the
temperature used there.

To give insight into some of the results just obtained, I consider their meaning for some simple models. For the
ideal intrinsic semiconductor~\cite{kittel,ashcroftmermin}
\begin{eqnarray}
E_1(N_0+1)-E_1(N_0)&=&\epsilon_c\label{H.30}\\
E_1(N_0+1)-E_1(N_0-1)&=&\epsilon_c+\epsilon_v\label{H.31}\\
E_1(N_0)-E_1(N_0-1)&=&\epsilon_v. \label{H.32}
\end{eqnarray}
If we define the gap in the many-body spectrum as $G = E_2(N_0)-E_1(N_0)$ then we get the familiar result,
$G=\epsilon_c-\epsilon_v.$ Remembering that the 1-electron energies are negative (the electrons are bound in the
solid), we see that $I\equiv E_1(N_0-1)-E_1(N_0)=-\epsilon_v>0$ as appropriate for the ionization energy,
similarly for the affinity. These are special cases of the general theorem ~\cite{kaplankleiner}
\begin{equation}
E_1(N+1)\le E_1(N),\label{monotone}
\end{equation}
valid beyond NEM's. Clearly $I-A=\epsilon_c-\epsilon_v = G$ within these NEM's. However, one should realize the
well-known fact that for an interacting system this relation between $G$ and $I-A$ does not hold in general. E.g.
consider the half-filled, zero-hopping, extended Hubbard model (including nearest-neighbor intersite Coulomb
interaction $V > 0$), with the number of sites (or electrons), $N_0\ge2$. For this, $I-A=U$ while $G = U-V$. $I-A$
is commonly called the ``band gap" for intrinsic semiconductors.

The important result~(\ref{H.26}), particularly for $\mathcal{N}= $integer,  applies to any finite ``molecule". It
is natural to ask if that result holds in general for a solid if the thermodynamic limit is taken before
$T\rightarrow 0$. I want to show that while the two limits commute in some familiar cases, they do not always do
so. I'll accomplish this by examples. Recollect first that the thermodynamic limit for a solid considers a
sequence of solids where atoms are added to the crystalline array keeping the volume per atom fixed (I will follow
the common approach using periodic boundary conditions). Generally then one can take the band edges as unchanged
in this sequence, while additional energies fill the space between. Consider the NEM of a semiconductor. For the
intrinsic case, and with the number of electrons = precisely the number of states in the valence band, the usual
result is the case where the TL is taken first, sums over wave vector having been converted to integrals, with the
zero-$T$ limit $\mu = (\epsilon_v+\epsilon_c)/2$; the opposite order of limits is the same, since the zero-T limit
is given by~(\ref{H.26}) which is the same, and the TL changes nothing. Similarly for the case of a semiconductor
with donor and acceptor impurity densities $n_d > n_a >0$ (assuming the impurity bands have zero width). Again the
usual result gives the zero-$T$ limit after the TL as~\cite{hannay} $\mu\rightarrow$ the donor level as
$T\rightarrow0$; and the zero-$T$ limit for the finite system, given by~(\ref{H.26}), is again the donor level,
which doesn't change under the TL. However, let's return to the ideal intrinsic case and add an electron. The
example in the Appendix shows that for $\mu(N_0+1)$ the two limits do not commute.

 As a final comment in this section, I give a
``physical" explanation of \emph{why} $\mu$ goes to the middle of the gap as $T\rightarrow0$ for an ideal
intrinsic semiconductor. This is based on the non-interacting model with the assumption that the rate of transfer
of an electron of energy $\epsilon_1$ from a semiconductor to an adjacent metal with c.p. $\mu_{met}$ is
proportional to $|\epsilon_1-\mu_{met}|n_e$, $n_e$ is the density of electrons in the semiconductor, with a
similar statement for transfer of a hole at energy $\epsilon_2$ from the semiconductor to the metal, with $n_e$
replaced by $n_h$. At very low $T$, there are some thermally excited electrons in the conduction band and holes in
the valence band, with $n_e=n_h$. Now bring a metal with c.p. $\mu_{met}$ into conducting contact with the
semi-conductor with c.p. $=\mu_{sc}$. In accordance with the understanding of the chemical potential as a quantity
whose gradient drives particle flow, the condition for no current is $\mu_{met}=\mu_{sc}$. Suppose $\mu_{sc}$ is
in the gap but above the middle of the gap. Then $|\epsilon_c-\mu_{met}|= \epsilon_c-\mu_{met} <
|\epsilon_v-\mu_{met}|=\mu_{met}-\epsilon_v$, producing a net flow of holes into the metal,  thus contradicting
the condition of no flow when the c.p.'s are equal. The only way for the electron and hole flow to be equal, and
hence to have no net flow, is to have $\epsilon_c-\mu_{met} = \mu_{met}-\epsilon_v$, from which one obtains the
desired result.
\section{Finite size effects on $\mu$.}
In light of Shegelski's finding of huge finite-size effects on $\phi$ even for samples as large as 1 cm$^3$, we
feel compelled to see if similar effects might occur with our definition of the c.p. I followed standard procedure
within the effective mass approximation, calculating $\mu$ via sums over wave vectors $\mathbf{k}$, the
thermodynamic limit being obtained when these sums are replaced by appropriate integrals. The sum was done
numerically for 1 and 2 dimensions; I show the results for the latter case, the other being similar.

The usual effective mass single-particle spectrum is
\begin{eqnarray}
\epsilon_k&=&-\hbar^2k^2/2m_h\equiv\epsilon_k^v\ \mbox{for}\ \ \epsilon_k<0\nonumber\\
&=&\epsilon_c+\hbar^2k^2/2m_e\equiv\epsilon_k^c\ \mbox{for}\ \epsilon_k>0,\label{H.41}
\end{eqnarray}
with $\mathbf{k}=2\pi/(La)(p_1\hat{x}_1+p_2\hat{x}_2), p_i=0,\pm1,\cdots,\pm(L-1)/2$. $La$ is the linear size of
the square lattice, $L$ being taken as odd. Using the grand canonical ensemble, the following is solved for $\mu$:
\begin{equation}
\sum_\mathbf{k}[f(\epsilon_k^v-\mu)+f(\epsilon_k^c-\mu)]=N_0,\label{H.42}
\end{equation}
where $N_0=L^2.$ I took $(2\pi^2\hbar^2)/(m_e a^2)$ = 10eV, which is roughly its value using the free electron
mass, $a=3\AA$ and a gap of 0.5eV ; also I assumed $m_h/m_e=2$. The results are given
\begin{table}[h]
\begin{tabular}{lllll}
$\hspace{.22in} L$&\hspace{.22in}9&\hspace{.18in}17&\hspace{.18in}35&\hspace{.17in}81\\
$\hspace{.1in}T$& & & &\\
.01&.25004&.25244&.25347&.25348\\
.005&.250005&.250294&.25171&.25173\\
.0025&.2500005&.250005&.25065&.25087\\
.00125& & &.2500&.25043\\
.001&&&.25003&.25034
\end{tabular}
\parbox{2.5in}{\caption[Table~\ref{Table:chemical potential}]{$\mu$ vs $T,L$; $\mu,T$ in eV.}
~\label{Table:chemical potential}}
\end{table}
\begin{figure}[h]
\centering\includegraphics[height=2.4in]{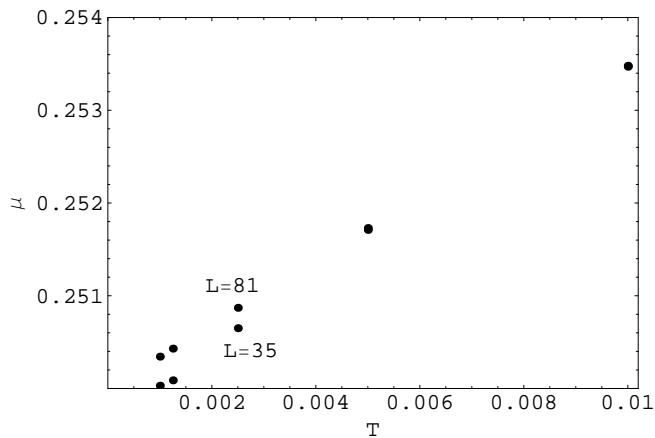}
 \caption{Chemical potential $\mu$ as function of $T$ and $L$. $\mu, T$ in eV.}\label{Fig:c.p.}
\end{figure}
in Table~\ref{Table:chemical potential} and in Figure~\ref{Fig:c.p.}, where just the two largest sizes are plotted
(I used the remarkable Arbitrary Precision command of Mathematica~\cite{wolfram}). For the largest $L$, the
results agree with the T. L. in 2-D
\begin{equation}
\mu_\infty=\frac{1}{2}(G+kTln\frac{m_h}{m_e}),\label{H.43}
\end{equation}
essentially perfectly down to $T=.001$, i.e. $T \sim 10K$. (Actually there are exponentially small corrections
to~(\ref{H.43}), which are negligible here.)

That this is for a system size of less than 7000 unit cells shows that our definition of the c.p. essentially
reaches the TL for systems far smaller than ``macroscopic".

\begin{figure}[h]
\centering\includegraphics[height=2.2in]{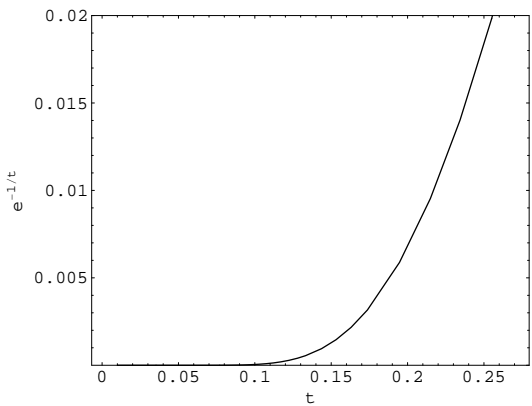}
 \caption{$e^{-1/t}$ vs $t$}\label{fig:exp}
\end{figure}

It is seen that finite size effects lead to a reduction in $\mu/\mu_\infty$ at low $T$. Further, the points for
$L=35$ appear to be heading to 0 at finite $T$ (they don't actually!). These facts can be understood as follows.
Basically the reduction occurs when $T$ becomes less than the spacing between levels near the band edges. The
seeming phase transition at $T\sim .001$ in the $L=35$ case arises because of the essential singularity at $T=0$
of $exp{(-T_0/T)}$. To see this, we plot $e^{-1/t}$ in Figure~\ref{fig:exp} which is seen to show a kind of
critical value $t_c$ such that the function appears to be zero for $t < t_c$, with $t_c\sim 0.1$. The level
spacing for the conduction band is $10/L^2 \sim .008$ for this case, which would give $T_c \sim .0008$, close to
the observed value .001 .

\section{Final remarks}
We have argued that the proper ensemble for considering the statistical thermodynamics of finite size systems is
the grand canonical ensemble whenever the chemical potential is physically relevant, and that the uniquely
correct definition $\mu$ of the chemical potential is the $\mu$ that is usually written in that ensemble's
density operator~($\nu$ in (\ref{H.4})). This definition satisfies the historical thermodynamic
definition~(\ref{thermomu}), the differentiation being defined because the average value of the number of
particles $\mathcal{N}$ in the grand canonical ensemble is continuous. This obviates the need for guesses such
as $\phi,\psi$ where one tries to define the c.p. as some functional of a free energy function defined only on
the integers, a need that occurs for the other standard ensembles. We note that this is a rather serious
departure from the usual view of statistical thermodynamics, where a fundamental equation obtained from any
ensemble is thought to give all fundamental equations. We discussed small-system experiments that support the
proposed definition, and showed that this definition avoided the ``bad" huge finite-size effects found by
Shegelski~\cite{shegelski1} for the ideal intrinsic semiconductor using $\phi$ as the definition of the chemical
potential.

It should be realized that since I have defined finite systems as systems with anywhere from a few to
$\sim10^{23}$ particles, for most `finite' systems on the large end of this range the difference between the
various ensembles will be negligible, the free energy will be extensive and the various quantities
$\mu,\phi,\psi$ will be essentially the same. It is for the smaller systems, and what may be considered the
anomalous ideal semiconductor model, that my conclusion is relevant. For a small system \emph{S} in contact with
a reservoir \emph{R}, one can of course consider the whole system, \emph{S} + \emph{R} as a new system. If
\emph{R} is macroscopic and not anomalous, the various ensembles and definitions of the c.p. will be equivalent
for \emph{S+R}. But this procedure entails consideration of the reservoir; the advantage of using a statistical
description of \emph{S} alone is that one can avoid consideration of reservoir details.  This point was
illustrated in the small-system examples presented above (the lead in the quantum dot example of course being
the reservoir).

Despite the fact that the definition $\phi$ of the c.p., used by many, is unacceptable, it is a well-defined
quantity. Shegelski's finding~\cite{shegelski1} that it shows huge finite size effects, should give one pause in
making the usual rather universal assumption that the TL of any quantity observable in principle is well
achieved for usual macroscopic systems. In particular, the argument that the definitions $\phi,\psi$ are
justified for macroscopic (and ideal intrinsic) semiconductors is proved false by this finding. Because I found
this phenomenon quite surprising, I give in the Appendix a relatively simple discussion that I believe makes it
plausible.

Interestingly, the result found there shows that, although using the \emph{correct} (i.e. the proposed)
definition of the c.p. avoided the bad behavior of $\phi$, similar ``bad" behavior is shown for this
\emph{correct} definition in the case of an added electron. It is important to note that this case is quite
special, the Fermi level $\mu$ sitting at a singularity in the density of states in the limit of zero
temperature. Whether or not surprising behavior will show up at other types of singularities, most macroscopic
systems will show a behavior of $\mu$ consistent with the usual expectation that it is very close to the
thermodynamic limit even at very low temperatures. E.g., in a metal, very low means $\stackrel{\sim}{>}$
separation of 1-electron energies (typically $\stackrel{\sim}{>} (10^5/N) K$, to within a few factors of 10
either way), tiny indeed, even in 2 dimensions where the number of atoms $N$ is around $10^{16}$. In this case
the distinction between $\phi,\psi,$ and $\mu$ essentially vanishes. Similarly, the presence of realistic
amounts of impurities in a semiconductor tends to remove the distinction between these quantities, as noted by
Shegelski~\cite{shegelski1,shegelski2} (surface states can act similarly). But for small systems the distinction
is unavoidable, leading me to consider the very-small-system examples discussed above as the physically most
important ones, although the ideal intrinsic semiconductor is of considerable pedagogical importance (it's where
one begins any study of semiconductors).

\textbf{Acknowledgments} I thank Jack Bass for bringing Shegelski's work to my attention and for many useful
discussions. I am indebted to S. D. (Bhanu) Mahanti for intensive discussions on the physics involved as well as
the general philosophy of the approach. I appreciate Phillip Duxbury's challenge concerning the quantum dot
example and Carlo Piermarocchi's pointing out the relevance of the book by Cohen-Tannoudji and other helpful
comments. I thank Stuart Tessmer for teaching me about the experimental physics of quantum dots and for many
helpful discussions, James Harrison for guiding me to the Mulliken papers and useful discussions, Simon Billinge
for a helpful discussion, Elliott Lieb for the Zhislin reference, and Mark Shegelski, Peter Landsberg, Neil
Ashcroft, and Herbert Kroemer for useful correspondence. I appreciate John Perdew's patience and clarity in
explaining his work and his encouragement.

\appendix
\section{Chemical potential for an electron added to an intrinsic semiconductor}
Working in the grand canonical ensemble, to calculate the term $\mathcal{F}(N_0+1)$ that appears in $\phi (N_0)$,
one requires the chemical potential $\mu(N_0+1)$. ($N_0$ is the number of electrons in the neutral semiconductor.)
$\mu(N_0+1)$ is simpler to calculate than $\mathcal{F}(N_0+1)$, and fortunately it shows huge finite size effects
like those found~\cite{shegelski1} for $\phi(N_0)$. I expect that this behavior of $\mu(N_0+1)$ is the origin of
the behavior of $\phi(N_0)$, and therefore present its calculation here with the purpose of making Shegelski's
result plausible to the skeptics (of which I was one). According to Shegelski, the source of the slow convergence
to the TL is a logarithmic dependence on the volume or $N_0$, so we will be looking for this.

I consider temperatures $>>$ the separation of levels near the bottom of the conduction band, in which case we can
replace sums over $\mathbf{k}$ by integrals. Writing
\begin{equation}
\mathcal{N}(\mu)=\frac{N_0}{\Delta}\int_0^\Delta f(y-\mu)dy + \frac{N_0}{\Delta'}\int_{\Delta+G}^\infty
f(y-\mu)dy,\label{A1}
\end{equation}
the equation determining $\mu(N_0+1)$ is then
\begin{equation}
\mathcal{N}(\mu(N_0+1))=N_0+1.\label{A2}
\end{equation}
In (\ref{A1}), $\Delta$ is the width of the valence band, $G$ is the gap, and $N_0/\Delta\equiv D,
N_0/\Delta'\equiv D'$ are the valence and conduction band densities of states (I'm considering dimensionality 2).
The integrals are elementary, giving
\begin{eqnarray}
\mathcal{N}(\mu)&=&N_0+DkTln\frac{1+e^{-\beta\mu}}{1+e^{-\beta(\mu-\Delta)}}\nonumber\\
& & +D'kT ln[1+e^{-\beta(\Delta+G-\mu)}].\label{A3}
\end{eqnarray}
We're interested in $kT<<G<\Delta$. As a check, if we equate this expression for $\mathcal{N}(\mu)$ to $N_0$ and
assume $\mu>\Delta$ we find (\ref{H.43}) for $\mu(N_0)$. (\ref{A2}) yields
\begin{equation}
\frac{1}{N_0}=\frac{kT}{\Delta'}ln[1+e^{-\beta(\Delta+G-\mu')}]-\frac{kT}{\Delta}ln[1+e^{-\beta(\mu'-\Delta)}],\label{A4}
\end{equation}
where $\mu(N_0+1)\equiv \mu'$. For simplicity, take $\Delta=\Delta'$. Then~(\ref{A4}) can be written as a
quadratic equation in $e^{\beta\mu'}$, which gives finally
\begin{equation}
\mu'=\Delta+G+kTln \frac{x-1+\sqrt{(x-1)^2+4x\  exp{(-\beta G)}}}{2},\label{A5}
\end{equation}
where $x=e^{\beta\Delta/N_0}$.

\begin{figure}[h]
\centering\includegraphics[height=1.7in]{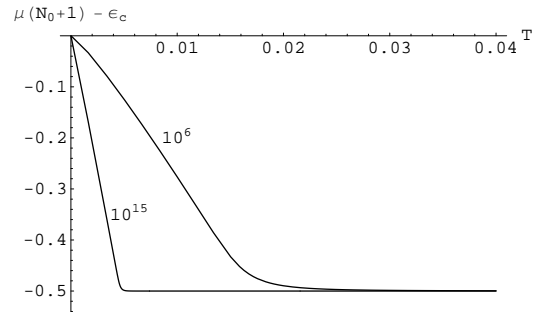} \vspace{.2in}
 \caption{Deviation of the chemical potential from the bottom of the conduction band,
 $\mu(N_0+1)-\epsilon_c$, in units of the gap, G=0.25eV, as a function of $T$ for $N_0 = 10^6$ and $10^{15}$.
 $T$ in eV.}\label{Fig:mu}
\end{figure}
(\ref{A5}) gives
\begin{equation}
\lim_{T\rightarrow0} \mu(N_0+1) = \Delta+G+\Delta/N_0,\label{A6}
\end{equation}
and
\begin{equation}
 \lim_{N_0\rightarrow\infty}\mu(N_0+1) = \Delta+G/2.\label{A7}
 \end{equation}
 The zero-$T$ limit is close to the bottom of the conduction
 band, as expected. This limit is questionable since we
 assumed $T >>$ the separation of conduction band levels, but see below for justification.
 The TL~(\ref{A7}) is where it must be, in the middle of the gap,
 because the difference in the numbers of electrons and holes is just 1 (in $N_0$).

 To look more closely at the approach to the TL, we note that $x-1\approx\beta\Delta/N_0$. Hence $\mu$ is
 analytic at $1/N_0=0$, according to~(\ref{A5}). I.e., there is no logarithmic $N_0$ dependence seen here. However, we
 note that the two terms under the square root compete at large $N_0$ and small $T$: At small enough $T$, fixed
 $N_0$, the term $(x-1)^2$ dominates, while at large enough $N_0$ the other term wins. Let $T_0$ be the temperature at
 which the two terms are equal. Putting
 \begin{eqnarray}
 z&=&\beta_0 G/2\label{A8}\\
 p&=&N_0G/(2\Delta),\label{A9}
 \end{eqnarray}
$T_0$ is determined by $e^{z/p}-e^{-z}=1$ or
\begin{equation}
z+ln\ z=ln\ p, \label{A10}
\end{equation}
for $p>>1$ and $z>> 1$. This says that $z\approx ln\ p$ as $p\rightarrow\infty$, and, further,~\cite{debruijn}
\begin{equation}
z=ln\ p-ln\ ln\ p + O(\frac{ln\ ln\ p}{ln\ p}). \label{A11}
\end{equation}
Thus, not only does $k T_0/G (\sim 1/z)$ approach 0 as $1/ln\ p$, but the corrections to this asymptotic behavior
become small very slowly. Thus we have seen just how logarithmic behavior enters.

For a quantitative illustration, we have plotted~(\ref{A5}) in Fig.~\ref{Fig:mu}, taking $G=0.25 eV, \Delta=10eV$.
It is seen that the TL, namely $\mu(N_0+1)$ being at the middle of the gap, is approached for high $T$, and moves
to the top of the gap at $T\rightarrow0$. It can be checked that the temperature $T_0$, below which deviation from
the TL begins to get big, tracks well with the first two terms of~(\ref{A11}). For the larger value $N_0=10^{15}$,
which is macroscopic (D=2), $T_0\sim 0.005eV \sim 50K$, so the TL is largely in error for $T \sim
25K$.~\cite{footnote4} This is very similar behavior to that found by Shegelski~\cite{shegelski1} for $\phi(N_0)$,
but I emphasize that the quantity $\mu(N_0+1)$ studied here is the true chemical potential for the
$N_0+1$-particle system, which is not $\phi(N_0)$. However the two quantities are related in that $\mu(N_0+1)$
enters directly the calculation of $\mathcal{F}(N_0+1)$, which enters directly into the definition of $\phi(N_0)$,
making it very plausible that the large finite size effects in $\phi$ (for macroscopic samples!) found by
Shegelski are real. One should remember that this $\phi$ is not the chemical potential for the neutral
semiconductor, as I have argued above. Nevertheless, it is well defined, and its large deviation from the
TL\cite{shegelski1} renders incorrect the widely
quoted~\cite{shegelski1,shegelski2,ashcroftmermin,landsberg,fetter,hill} $\phi$, and
similarly~\cite{kittel2,fetter} $\psi$ (I presume), as representing the chemical potential for this neutral
system.



\thebibliography{0}
\bibitem{footnote1} But later in his papers these quantities may be calculated either in the canonical or in the
grand canonical ensemble.
\bibitem{footnote2}This was called an ideal intrinsic semiconductor~\cite{shegelski1}; it is the usual textbook
model where noninteracting electrons move in a periodic potential, treated by periodic boundary conditions (there
are neither impurities nor surface states).
\bibitem{footnoteTL} The condition $T>0$ must be understood: the TL of $\phi$
can have a discontinuity at $T=0$, as in the semiconductor model just discussed. I.e., $\phi$ (and $\psi$
of~(\ref{H.3}) can depend on the order in which the TL and the limit $T\rightarrow0$ are taken.
\bibitem{footnoteperdew} Non-integral average values of the number of particles were also considered
in~\cite{perdew,perdew2}.
\bibitem{footnotelocalspins} Magnetic properties of systems of localized spins are examples where the chemical
potential may not be physically relevant.
\bibitem{footnote3} Shegelski~\cite{shegelski1} has argued that the canonical ensemble is most appropriate
for a semiconductor, since it has a fixed number of electrons. I fault this argument by noting that, physically,
fluctuations in the number of electrons will occur in semiconductors as well as in metals. This is particularly
true for the situation where the semiconductor is put in contact with a piece of metal in order to consider
electrical current. Furthermore, these (relative) fluctuations for the grand canonical ensemble within the
noninteracting effective mass model can be seen to be negligible for a macroscopic semiconductor.
\bibitem{footnotegeneralization}The formal generalization of~(\ref{H.18}) and~(\ref{thermomu}) to a multi-component
system is straightforward. One has $N_i$ particles of type $i$, $\mu \hat{N}$ in~(\ref{H.13}) is replaced by $\sum
\mu_i \hat{N}_i$, (\ref{thermomu}) is replaced by $\partial\mathcal{F}/\partial\mathcal{N}_i=\mu_i$,
$\mathcal{N}_i$ is the expectation value of $N_i$, and the derivative is taken holding the other $\mathcal{N}_j$
constant.
\bibitem{footnotebroadening} There is another source of peak width in addition to the thermal effect considered here,
namely that due to the hopping of electrons between the levels of the dot and the overlapping continuum of the
leads. Under certain conditions the thermal broadening can be the dominant effect. See~\cite{kastner}.
\bibitem{footnotelandsbergtheorem}This is equivalent to the theorem of reference\cite{landsberg} for NEM's.
\bibitem{footnote4} These temperatures are high enough to justify the replacement of the sums by integrals.
\endthebibliography
\end{document}